\documentclass[twocolumn,amsmath,amssymb,10pt,superscriptaddress,a4paper,letterpaper,fleqn]{revtex4-1}
\usepackage{amssymb}
\usepackage{epsfig}
\usepackage{graphicx}
\usepackage{dcolumn}
\usepackage{array}
\usepackage{bm}
\usepackage{fancyheadings}
\usepackage{longtable}
\usepackage{multirow}
\usepackage{float}
\usepackage{afterpage}  
\usepackage{color}
\setlongtables
\usepackage[breaklinks=true,linkbordercolor={1 1 1}]{hyperref}
\begin{document}
\setcounter{page}{1}
\title{
     \qquad \\ \qquad \\ \qquad \\  \qquad \\  \qquad \\ \qquad \\ 
 Calculations of Nuclear Astrophysics and Californium Fission Neutron Spectrum Averaged Cross Section Uncertainties using ENDF/B-VII.1, JEFF-3.1.2, JENDL-4.0 and Low-Fidelity Covariances
}

\author{Boris Pritychenko}
   \email[Corresponding author: ]{pritychenko@bnl.gov}
   \affiliation{National Nuclear Data Center, Brookhaven National Laboratory, Upton, NY 11973, USA}

\date{\today}
   \received{XX May 2014; revised received XX August 2014; accepted XX September 2014}

\begin{abstract}{
Nuclear astrophysics and californium fission neutron spectrum averaged  cross sections  
and their uncertainties for ENDF materials have been calculated. Absolute values were deduced with  
Maxwellian and Mannhart spectra, while uncertainties are based on ENDF/B-VII.1, JEFF-3.1.2, JENDL-4.0 and Low-Fidelity covariances. 
These quantities are compared with available data, independent benchmarks, EXFOR library, 
and analyzed for a wide range of cases.  Recommendations for neutron cross section covariances are given 
and implications are discussed.
}
\end{abstract}
\maketitle

\lhead{Calculations of Nuclear Astrophysics  $\dots$}    
\chead{NUCLEAR DATA SHEETS}                       
\rhead{Boris Pritychenko}              
\lfoot{}                                                           
\rfoot{}                                                          
\renewcommand{\footrulewidth}{0.4pt}


\section{ INTRODUCTION}

Calculations of integral values at NNDC have been conducted in parallel with the ENDF/B-VII library 
releases \cite{2006Chad,2011Chad}. These values  represent the complementary data sets for nuclear 
astrophysics, industry, and data evaluation   applications.  First results on reaction rates and neutron 
cross sections \cite{2010Pri} have demonstrated a large  potential of ENDF/B-VII for applications, 
such as  KADoNiS stellar nucleosynthesis library \cite{2006Dil}.  Further interactions with the fundamental and 
applied science communities have initiated work on the extended list  of integral values and 
their uncertainties \cite{2012Pri,2006Mugh,2007Iaea}. Calculations of nuclear astrophysics and californium fission neutron spectrum 
averaged cross section ({\it i.e.} californium spectrum)  uncertainties are presented in the following sections.

\section{Maxwellian-Averaged Cross Sections Uncertainties}

\begin{figure}[!htb]
  \includegraphics[width=0.75\columnwidth]{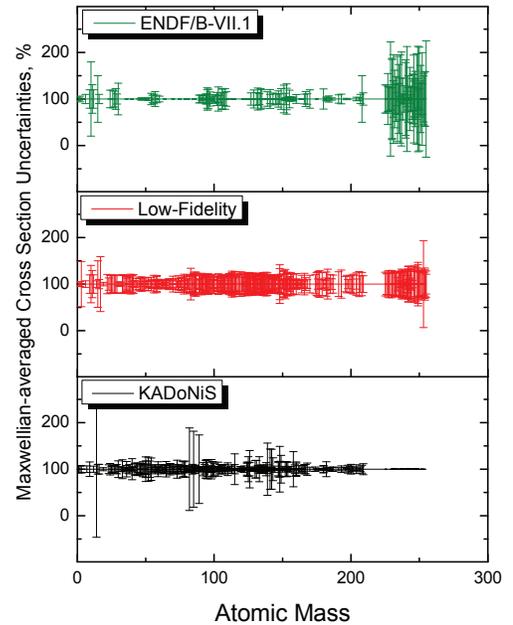}
  \caption{(Color online) Maxwellian-averaged neutron capture cross section, {\it kT}=30 keV, uncertainties for ENDF/B-VII.1 library, Low-Fidelity project and KADoNiS database \cite{2011Chad,2008Lit,2006Dil}. Data are taken from \cite{2012Pri}.}
  \label{fig:MaxUn}
\end{figure}

Nuclear data covariances are essential for fundamental and applied nuclear science and technology. 
They provide the experimentally-observable uncertainties  that are necessary for application development. 
Maxwellian-averaged cross sections and their uncertainties have been calculated in recent years \cite{2010Pri,2012Pri}. 
 Fig. \ref{fig:MaxUn} shows cross section uncertainties for ENDF/B-VII.1 evaluated library, Low Fidelity project, and KADoNiS database \cite{2011Chad,2008Lit,2006Dil} demonstrate nuclear astrophysics value of  
 ENDF and Low Fidelity covariances  for stellar nucleosynthesis research. At the same time, the  ENDF/B-VII.1 and Low Fidelity 
uncertainties are relatively large for precise calculations. The stellar nucleosynthesis calculations  require the 
stringent cross section uncertainties in order $<$3$\%$ to finalize the branching of the {\it s}-process path. However, even a 
specially-designated KADoNiS library, at the present state, cannot satisfy this requirement, and further research is necessary.

\section{$^{252}$Cf Fission Neutron Spectrum Averaged Cross Sections and their Uncertainties}

\begin{figure}[!htb]
  \includegraphics[width=0.75\columnwidth]{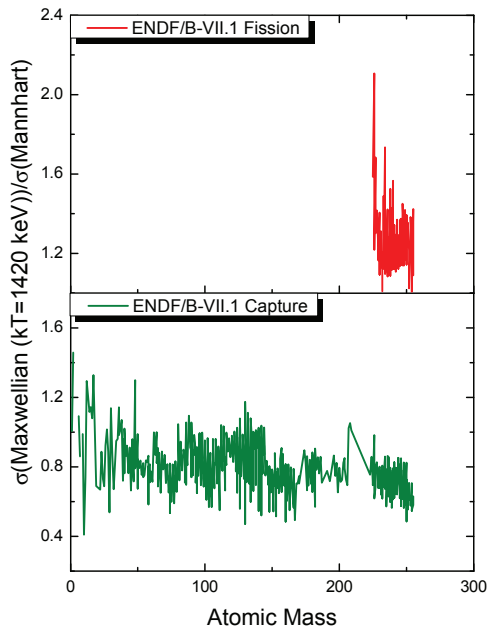}
  \caption{(Color online) The ratio of calculated ENDF/B-VII.1 californium spectrum neutron cross sections using Maxwellian, {\it kT}=1420 keV, and Mannhart spectra \cite{2012Pri,2002Mann}.}
  \label{fig:KTMann}
\end{figure}
$^{252}$Cf is often used in nuclear physics as a compact, portable and intense
neutron source. 
Its  neutron energy spectrum is similar to a fission reactor, with an average energy of 2.13 MeV. 
This is very convenient for ENDF libraries validation tests in the fast region, even though 
it is not exactly representative of a fast reactor spectrum (being hotter) \cite{2014Chad}.

For  evaluation purposes, 
$^{252}$Cf spectrum neutron fission and capture averaged cross sections were calculated using Maxwellian-averaged ({\it kT}=1420 keV) spectrum and Mannhart evaluation \cite{2012Pri,2002Mann}.  Fig. \ref{fig:KTMann} shows the ratio of calculated californium spectra cross sections using Maxwellian, and Mannhart approaches for ENDF/B-VII.1 library. 
This ratio indicates that Maxwellian spectrum provides a reasonable fit of californium data, however, it falls short 
of being used for nuclear standards and dosimetry purposes. Consequently, the Mannhart evaluation has been chosen for 
calculation of californium spectrum cross sections. 

\begin{figure}[!htb]
  \includegraphics[width=0.75\columnwidth]{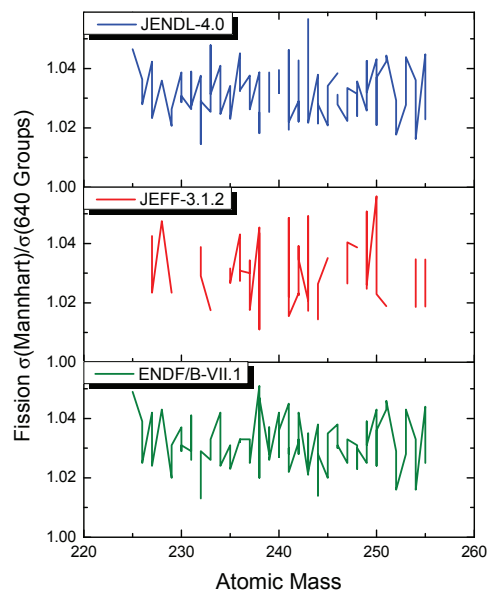}
  \caption{(Color online) The ratio of ENDF, JEFF, and JENDL calculated californium spectrum neutron fission cross sections using the original and 640-group Mannhart spectra \cite{2002Mann}.}
  \label{fig:FissionRatio}
\end{figure}

\begin{figure}[!htb]
  \includegraphics[width=0.75\columnwidth]{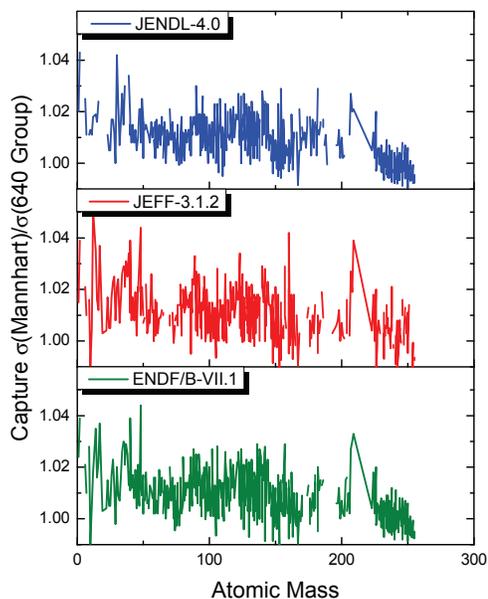}
  \caption{(Color online) The ratio of ENDF, JEFF, and JENDL  calculated californium spectrum neutron capture cross sections using the original and 640-group Mannhart spectra \cite{2002Mann}.}
  \label{fig:CaptureRatio}
\end{figure}
Presently, the original and 640-group representations of Mannhart evaluation are frequently considered. To evaluate a possible spectrum representation impact,  neutron cross sections have been calculated  using the both formats. 
Figs. \ref{fig:FissionRatio}, \ref{fig:CaptureRatio} show the ratios of californium fission and capture cross sections for both representations of the  
 linearized ENDF libraries.  The plotted ratios  clearly demonstrate the impact of different representations.

Following the nuclear dosimetry example,  $^{252}$Cf spectrum neutron fission averaged cross sections for major evaluated libraries: ENDF/B-VII.1, JEFF-3.1.2, and JENDL-4.0 
 \cite{2011Chad,2011Kon,2011Shi} have been produced using the 640-group format and shown in the Table \ref{FissionTable}.  
 These data are in a good agreement with the previously-published CIELO values \cite{2014Chad}.  Californium  spectrum neutron capture averaged cross sections are available upon request.

\renewcommand{\arraystretch}{0.5}
\renewcommand\tabcolsep{1pt}
\scriptsize
\begin{center}

\begin{longtable*}{@{}>{\small}l|>{\small}c>{\small}c>{\small}c>{\small}c@{}}
\caption{640-group californium spectrum  neutron fission averaged  cross sections for ENDF, JEFF, and JENDL major evaluated libraries, and EXFOR (experimental nuclear reaction) data ({\it kT}$\sim$1.42 MeV) \cite{2011Chad,2011Kon,2011Shi,2014Otu}.} \label{FissionTable}
 \\
\hline \hline \\
\textbf{  Material  } & \textbf{  ENDF/B-VII.1  } & \textbf{  JEFF-3.1.2  } & \textbf{  JENDL-4.0  } & \textbf{  EXFOR}  \\
   	& (barns) & (barns) & (barns) & (barns) 	\\
\hline \\
\endfirsthead
\caption[]{640-group californium spectrum neutron  $\dots$  (continued).}
\\
\hline \hline \\
\textbf{Material} & \textbf{ENDF/B-VII.1} & \textbf{JEFF-3.1.2} & \textbf{JENDL-4.0} &  \textbf{EXFOR}  \\
   	& (barns) & (barns) & (barns) & (barns) 	\\
\hline \\
\endhead

 88-Ra-223 & 5.485E-2$\pm$8.293E-4 & 5.485E-2$\pm$8.293E-4 & 5.485E-2$\pm$8.293E-4 &  \\
 88-Ra-224 &     &     &     &  \\
 88-Ra-225 &     &     &     &  \\
 88-Ra-226 & 3.741E-4$\pm$5.990E-6 & 3.741E-4$\pm$5.990E-6 & 3.740E-4$\pm$5.988E-6 &  \\
 89-Ac-225 & 3.505E-2$\pm$4.369E-2 &     & 3.503E-2$\pm$4.379E-2 &  \\
 89-Ac-226 & 3.478E-2$\pm$4.126E-2 &     & 3.472E-2$\pm$4.337E-2 &  \\
 89-Ac-227 & 1.253E-2$\pm$2.089E-3 & 1.316E-2$\pm$1.755E-4 & 1.252E-2$\pm$2.082E-3 &  \\
 90-Th-227 & 2.968E-1$\pm$2.358E-1 & 4.747E-1$\pm$7.820E-3 & 2.966E-1$\pm$2.402E-1 &  \\
 90-Th-228 & 3.768E-1$\pm$7.536E-2 & 1.073E-1$\pm$1.443E-3 & 3.757E-1$\pm$7.561E-2 &  \\
 90-Th-229 & 3.433E-1$\pm$5.660E-2 & 4.747E-1$\pm$7.819E-3 & 3.431E-1$\pm$5.780E-2 &  \\
 90-Th-230 & 2.044E-1$\pm$1.592E-2 &     & 2.042E-1$\pm$1.594E-2 &  \\
 90-Th-231 & 1.977E-1$\pm$1.630E-1 &   & 1.977E-1$\pm$1.623E-1 &  \\
 90-Th-232 & 7.582E-2$\pm$1.824E-3 &     & 8.170E-2$\pm$6.032E-3 & 8.470E-2$\pm$4.900E-3 \\
 90-Th-233 & 9.916E-2$\pm$9.237E-2 & 1.084E-1$\pm$1.850E-3 & 9.918E-2$\pm$9.174E-2 &  \\
 90-Th-234 & 3.542E-2$\pm$4.487E-2 &     & 3.541E-2$\pm$4.396E-2 &  \\
 91-Pa-229 & 1.939E+0$\pm$4.916E-1 &   & 1.938E+0$\pm$4.862E-1 &  \\
 91-Pa-230 & 1.782E+0$\pm$3.484E-1 &   & 1.781E+0$\pm$3.463E-1 &  \\
 91-Pa-231 & 7.667E-1$\pm$1.031E-2 & 9.843E-1$\pm$1.330E-2 & 8.442E-1$\pm$2.385E-2 & 9.700E-1$\pm$4.500E-2 \\
 91-Pa-232 & 9.581E-1$\pm$2.711E-1 & 1.082E+0$\pm$1.739E-2 & 9.572E-1$\pm$2.804E-1 &  \\
 91-Pa-233 & 2.384E-1$\pm$3.065E-3 &     & 2.463E-1$\pm$5.697E-2 &  \\
 92-U -230 & 2.377E+0$\pm$5.504E-1 &   & 2.375E+0$\pm$5.416E-1 &  \\
 92-U -231 & 2.162E+0$\pm$3.938E-1 &   & 2.161E+0$\pm$3.900E-1 &  \\
 92-U -232 & 2.038E+0$\pm$6.916E-2 & 2.442E+0$\pm$3.590E-2 & 2.038E+0$\pm$6.916E-2 &  \\
 92-U -233 & 1.867E+0$\pm$3.478E-2 & 1.883E+0$\pm$3.130E-2 & 1.879E+0$\pm$2.949E-2 & 1.947E+0$\pm$3.100E-2 \\
 92-U -234 & 1.186E+0$\pm$2.199E-1 & 1.171E+0$\pm$1.591E-2 & 1.211E+0$\pm$3.167E-2 &  \\
 92-U -235 & 1.209E+0$\pm$2.000E-2 & 1.203E+0$\pm$1.913E-2 & 1.202E+0$\pm$2.162E-2 & 1.266E+0$\pm$1.823E-2 \\
 92-U -236 & 5.873E-1$\pm$1.371E-1 & 6.059E-1$\pm$7.922E-3 & 5.801E-1$\pm$9.429E-3 &  \\
 92-U -237 & 6.320E-1$\pm$9.515E-3 & 8.719E-1$\pm$1.339E-2 & 5.874E-1$\pm$8.340E-2 &  \\
 92-U -238 & 3.117E-1$\pm$4.753E-3 & 3.102E-1$\pm$4.006E-3 & 3.094E-1$\pm$4.815E-3 &  3.109E-1$\pm$1.400E-2 \\
 92-U -239 & 3.730E-1$\pm$5.929E-3 &   &   &  \\
 92-U -240 & 1.953E-1$\pm$2.549E-3 &   &   &  \\
 92-U -241 & 1.881E-1$\pm$2.812E-3 &   &   &  \\
 93-Np-234 & 2.436E+0$\pm$4.196E-1 &   & 2.436E+0$\pm$4.243E-1 &  \\
 93-Np-235 & 2.173E+0$\pm$5.521E-1 & 1.878E+0$\pm$2.709E-2 & 2.173E+0$\pm$5.495E-1 &  \\
 93-Np-236 & 2.062E+0$\pm$3.925E-1 & 1.891E+0$\pm$2.975E-2 & 2.062E+0$\pm$3.888E-1 &  \\
 93-Np-237 & 1.339E+0$\pm$4.624E-2 & 1.313E+0$\pm$1.773E-2 & 1.322E+0$\pm$2.803E-2 & 1.442E+0$\pm$2.300E-2 \\
 93-Np-238 & 1.431E+0$\pm$1.659E-1 & 1.453E+0$\pm$2.560E-2 & 1.430E+0$\pm$1.645E-1 &  \\
 93-Np-239 & 5.923E-1$\pm$5.151E-1 &     & 5.916E-1$\pm$5.341E-1 &  \\
 94-Pu-236 & 2.324E+0$\pm$2.389E-1 & 2.075E+0$\pm$3.180E-2 & 2.324E+0$\pm$2.390E-1 &  \\
 94-Pu-237 & 2.399E+0$\pm$6.080E-1 & 2.954E+0$\pm$4.526E-2 & 2.400E+0$\pm$6.024E-1 &  \\
 94-Pu-238 & 1.925E+0$\pm$4.890E-2 & 1.973E+0$\pm$2.816E-2 & 1.944E+0$\pm$7.337E-2 &  \\
 94-Pu-239 & 1.774E+0$\pm$2.865E-2 & 1.774E+0$\pm$2.653E-2 & 1.777E+0$\pm$2.654E-2 & 1.947E+0$\pm$3.100E-2 \\
 94-Pu-240 & 1.334E+0$\pm$1.993E-2 & 1.353E+0$\pm$1.822E-2 & 1.316E+0$\pm$1.779E-2 &  \\
 94-Pu-241 & 1.579E+0$\pm$3.618E-2 & 1.631E+0$\pm$2.579E-2 & 1.601E+0$\pm$3.585E-2 &  \\
 94-Pu-242 & 1.140E+0$\pm$2.945E-2 & 1.162E+0$\pm$1.560E-2 & 1.140E+0$\pm$2.932E-2 &  \\
 94-Pu-243 & 1.062E+0$\pm$1.501E-2 & 1.062E+0$\pm$1.501E-2 &   &  \\
 94-Pu-244 & 1.028E+0$\pm$2.802E-2 &     & 1.029E+0$\pm$2.792E-2 &  \\
 94-Pu-246 & 5.787E-1$\pm$5.108E-1 &     & 5.779E-1$\pm$5.082E-1 &  \\
 95-Am-240 & 1.987E+0$\pm$3.776E-1 &   & 1.986E+0$\pm$3.734E-1 &  \\
 95-Am-241 & 1.362E+0$\pm$3.067E-2 & 1.377E+0$\pm$1.805E-2 & 1.385E+0$\pm$3.027E-2 &  \\
 95-Am-242 & 1.950E+0$\pm$3.116E-2 & 1.726E+0$\pm$2.811E-2 & 1.807E+0$\pm$1.844E-1 &  \\
 95-Am-242M & 1.903E+0$\pm$3.419E-1 & 1.813E+0$\pm$2.934E-2 & 1.807E+0$\pm$5.952E-2 & 1.600E+0$\pm$2.200E-1 \\
 95-Am-243 & 1.075E+0$\pm$1.197E-1 & 1.077E+0$\pm$1.407E-2 & 1.076E+0$\pm$3.698E-2 & 1.145E+0$\pm$2.300E-2 \\
 95-Am-244 & 1.735E+0$\pm$2.843E-2 & 1.735E+0$\pm$2.843E-2 & 1.317E+0$\pm$3.417E-1 &  \\
 95-Am-244M & 1.735E+0$\pm$2.843E-2 & 1.735E+0$\pm$2.843E-2 & 1.162E+0$\pm$3.674E-1 &  \\
 96-Cm-240 & 2.026E+0$\pm$4.665E-1 & 1.719E+0$\pm$2.236E-2 & 2.025E+0$\pm$4.587E-1 &  \\
 96-Cm-241 & 2.184E+0$\pm$5.146E-1 & 2.639E+0$\pm$3.999E-2 & 2.182E+0$\pm$5.178E-1 &  \\
 96-Cm-242 & 1.755E+0$\pm$2.219E-1 & 1.643E+0$\pm$2.265E-2 & 1.754E+0$\pm$2.272E-1 &  \\
 96-Cm-243 & 2.396E+0$\pm$8.415E-2 & 2.137E+0$\pm$3.385E-2 & 2.396E+0$\pm$8.418E-2 &  \\
 96-Cm-244 & 1.717E+0$\pm$7.219E-2 & 1.591E+0$\pm$2.170E-2 & 1.717E+0$\pm$7.195E-2 &  \\
 96-Cm-245 & 1.730E+0$\pm$7.167E-2 & 1.711E+0$\pm$2.718E-2 & 1.730E+0$\pm$7.159E-2 &  \\
 96-Cm-246 & 1.240E+0$\pm$7.220E-2 & 1.217E+0$\pm$1.615E-2 &     &  \\
 96-Cm-247 & 1.835E+0$\pm$4.952E-2 & 1.886E+0$\pm$2.894E-2 & 1.835E+0$\pm$4.955E-2 &  \\
 96-Cm-248 & 1.081E+0$\pm$8.008E-2 & 1.240E+0$\pm$1.669E-2 & 1.081E+0$\pm$7.968E-2 &  \\
 96-Cm-249 & 1.202E+0$\pm$5.710E-1 & 2.065E+0$\pm$3.173E-2 & 1.202E+0$\pm$5.847E-1 &  \\
 96-Cm-250 & 6.630E-1$\pm$4.282E-1 & 1.538E+0$\pm$1.999E-2 & 6.630E-1$\pm$4.383E-1 &  \\
 97-Bk-245 & 1.087E+0$\pm$2.234E-1 &   & 1.084E+0$\pm$2.286E-1 &  \\
 97-Bk-246 & 1.732E+0$\pm$3.825E-1 &   & 1.732E+0$\pm$3.889E-1 &  \\
 97-Bk-247 & 8.941E-1$\pm$3.688E-1 & 1.014E+0$\pm$1.352E-2 & 8.932E-1$\pm$3.681E-1 &  \\
 97-Bk-248 & 1.541E+0$\pm$3.586E-1 &   & 1.541E+0$\pm$3.563E-1 &  \\
 97-Bk-249 & 1.035E+0$\pm$1.564E-1 & 9.726E-1$\pm$1.266E-2 & 1.033E+0$\pm$1.559E-1 &  \\
 97-Bk-250 & 1.003E+0$\pm$7.197E-1 & 2.013E+0$\pm$3.048E-2 & 1.003E+0$\pm$7.174E-1 &  \\
 98-Cf-246 & 2.152E+0$\pm$6.765E-1 &   & 2.153E+0$\pm$6.627E-1 &  \\
 98-Cf-248 & 1.323E+0$\pm$5.496E-1 &   & 1.318E+0$\pm$6.115E-1 &  \\
 98-Cf-249 & 1.723E+0$\pm$6.529E-2 & 1.723E+0$\pm$2.679E-2 & 1.721E+0$\pm$6.506E-2 &  \\
 98-Cf-250 & 1.489E+0$\pm$8.122E-1 & 1.877E+0$\pm$2.516E-2 & 1.488E+0$\pm$8.109E-1 &  \\
 98-Cf-251 & 1.273E+0$\pm$4.892E-1 & 1.696E+0$\pm$2.646E-2 & 1.273E+0$\pm$4.870E-1 &  \\
 98-Cf-252 & 2.291E+0$\pm$1.048E-1 & 1.897E+0$\pm$2.562E-2 & 2.291E+0$\pm$1.047E-1 &  \\
 98-Cf-253 & 7.674E-1$\pm$3.941E-1 &   & 7.677E-1$\pm$3.918E-1 &  \\
 98-Cf-254 & 1.779E+0$\pm$6.068E-1 & 2.133E+0$\pm$3.024E-2 & 1.773E+0$\pm$6.573E-1 &  \\
 99-Es-251 & 1.355E+0$\pm$7.800E-1 &   & 1.352E+0$\pm$7.677E-1 &  \\
 99-Es-252 & 2.143E+0$\pm$6.295E-1 &   & 2.143E+0$\pm$6.307E-1 &  \\
 99-Es-253 & 1.028E+0$\pm$7.579E-1 &     & 1.027E+0$\pm$7.546E-1 &  \\
 99-Es-254 & 1.901E+0$\pm$1.745E-1 & 2.153E+0$\pm$3.260E-2 & 1.899E+0$\pm$1.766E-1 &  \\
 99-Es-254M & 1.885E+0$\pm$2.099E-1 &   & 1.884E+0$\pm$2.076E-1 &  \\
 99-Es-255 & 7.059E-1$\pm$6.022E-1 & 2.222E+0$\pm$3.150E-2 & 7.053E-1$\pm$5.969E-1 &  \\
 100-Fm-255 & 2.189E+0$\pm$6.989E-1 & 2.294E+0$\pm$3.473E-2 & 2.188E+0$\pm$7.099E-1 &  \\
   \\
   \hline \hline
  \end{longtable*}

  \end{center}
\normalsize

\section{Cross Section Uncertainties Analysis and Recommendations}

To evaluate ENDF libraries covariances in the fast neutrons region I will consider Maxwellian, and  californium spectra  cross sections uncertainties, 
 and deduce recommendations. 
\begin{figure}[!htb]
  \includegraphics[width=0.75\columnwidth]{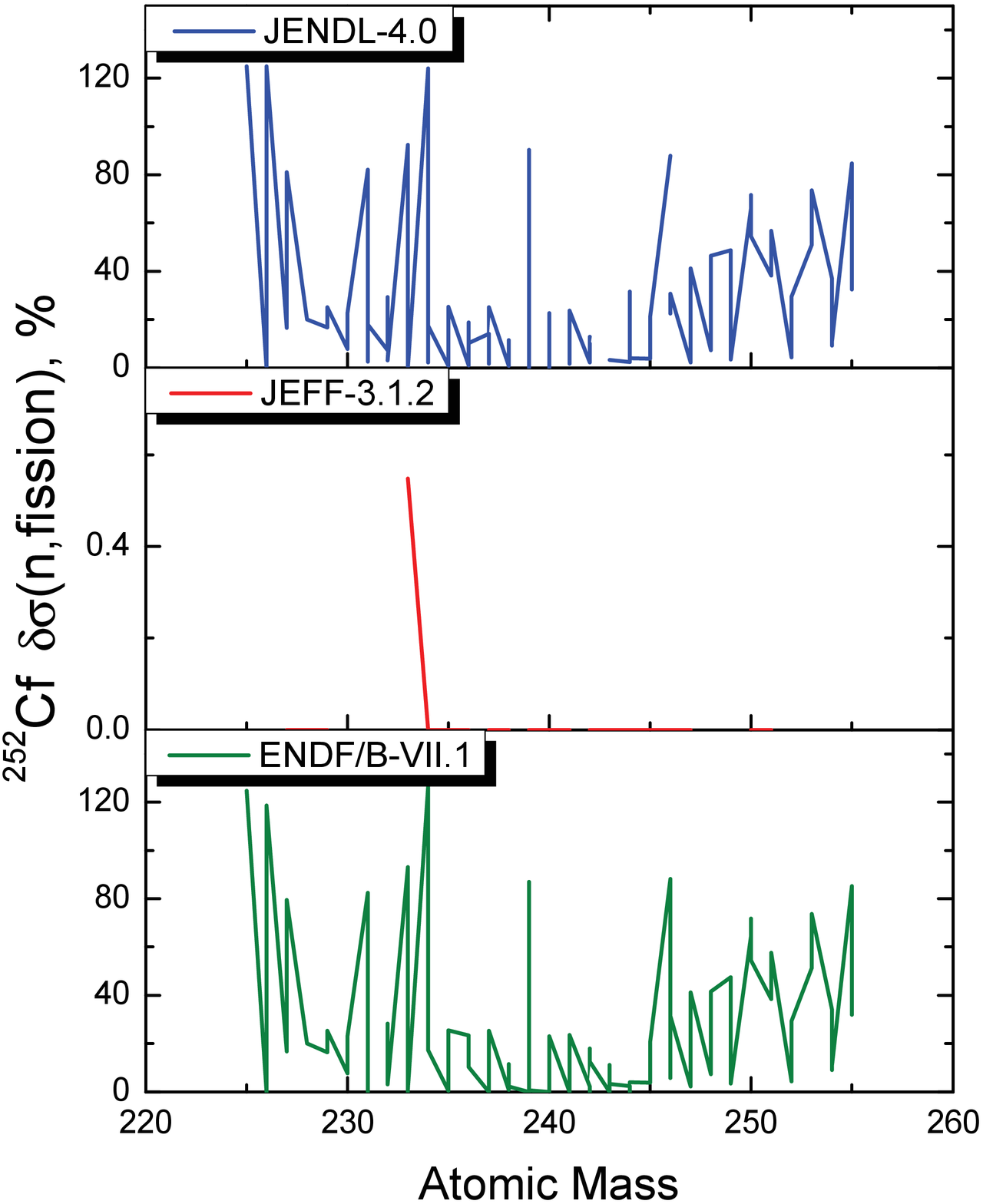}
  \caption{(Color online) The ENDF, JEFF, and JENDL calculated californium spectrum neutron fission cross section  uncertainties using the 640-group Mannhart spectrum \cite{2002Mann}.}
  \label{fig:FissionCov}
\end{figure}
\begin{figure}[!htb]
  \includegraphics[width=0.75\columnwidth]{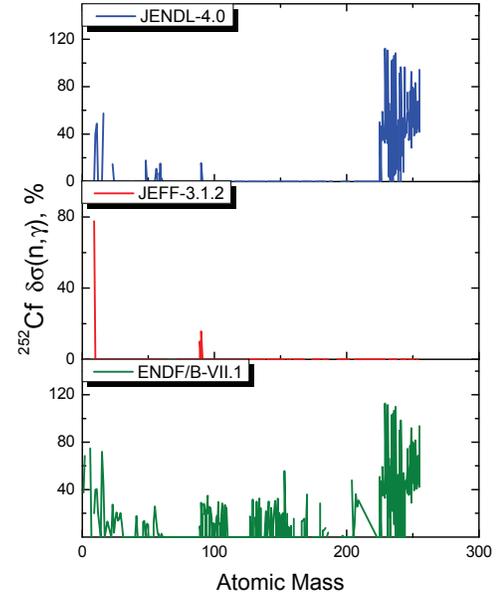}
  \caption{(Color online) The ENDF, JEFF, and JENDL  calculated californium spectrum neutron capture cross section uncertainties using the 640-group Mannhart spectrum \cite{2002Mann}.}
  \label{fig:CaptureCov}
\end{figure}

Visual inspection of the shown in Figs. \ref{fig:FissionCov}, \ref{fig:CaptureCov} data allows to spot the ``suspect" cases, where uncertainties are not very useful for application development. 
These somewhat unrealistic uncertainties of above 100 and below 1 $\%$ originate from theoretical models and fitting procedures, respectively. 
The summary of re-analysis of the previous Maxwellian data ({\it kT}=30 keV) \cite{2012Pri} and analysis of the current Mannhart spectrum uncertainties 
for ENDF/B-VII.1 library is shown in the Table \ref{tab:uncert}.
  \begin{table*}
    \caption{The summary of the ENDF/B-VII.1 library cross section uncertainties analysis.}\label{tab:uncert}
    \centering
    \begin{tabular}{p{1.5cm}|p{3cm}|p{3cm}|p{3cm}|p{3cm}}\hline\hline
               & \multicolumn{2}{c|}{} & \multicolumn{2}{c}{}  \\
      Reaction & \multicolumn{2}{c|}{Maxwellian spectrum, {\it kT}=30 keV} & \multicolumn{2}{c}{Mannhart spectrum \cite{2002Mann}}  \\\cline{2-5}
           &  &  &  &  \\  
           & Uncertainty $<$1$\%$ & Uncertainty  $>$100$\%$ & Uncertainty  $<$1$\%$ & Uncertainty  $>$100$\%$ \\\hline
     	&  &  &  & \\
     (n,fission) & $^{235}$U, $^{239,240}$Pu & $^{225,226}$Ac, $^{233}$Th, $^{229}$Pa, $^{235}$Np, $^{246}$Pu, $^{250}$Cm, $^{247,250}$Bk, $^{246,248,250,254}$Cf, $^{251,253,255}$Es & $^{235,238}$U, $^{239,240}$Pu & $^{225,226}$Ac, $^{234}$Th \\\hline
        &  &  &  & \\
     (n,$\gamma$) &  & $^{229}$Pa, $^{237}$Pu, $^{249}$Cm, $^{250}$Bk, $^{255}$Fm & $^{52}$Cr & $^{229}$Pa, $^{231}$U, $^{234,235,236}$Np, $^{237}$Pu \\\hline\hline
      
    \end{tabular}
  \end{table*}

The present analysis suggests the following recommendations for ENDF integral values and covariances:
\begin{enumerate}
 \item Absolute cross section values for linearized files are sensitive to the changes of Mannhart evaluation group structure. Calculated values are model dependent and may vary within 1-5$\%$.
 \item Nuclear astrophysics and energy applications require covariances for all ENDF materials.
 \item Realistic covariances are needed:
 \begin{itemize}
 \item Covariance matrices that result in $>$100 \% cross section uncertainties should be avoided, such large uncertainties are not very useful for application development.
 \item Covariance matrices that result in $<$1 \% cross section uncertainties are not realistic; strong contradiction with the best experiments.
 \item Presently, covariance matrices produce wide variations of cross section uncertainties within 0.5-120 \% range. This spread should be kept within 3-50 \% range.
 \end{itemize}
 \item Multiple MF=33 covariance matrices can be confusing.
\end{enumerate}

\section{CONCLUSIONS}

The previously-calculated ENDF/B-VII.1 and Low-Fidelity Maxwellian-averaged cross section uncertainties  have been re-analyzed.  
Californium spectrum neutron fission   and capture  averaged cross sections and their uncertainties have been calculated for 
ENDF/B-VII.1, JEFF-3.1.2, and JENDL-4.0 nuclear data libraries.  Recommendations for ENDF covariances 
have been deduced using the application development needs.\\


{\it Acknowledgements:} The author thanks M. Herman (BNL) for support of this work, R. Capote, A. Trkov and V. Zerkin (IAEA) for help with Mannhart spectra and data processing, and   
Mrs. M. Blennau (BNL)  for careful reading of the manuscript and useful suggestions. 
Work at BNL was funded by the Office of Nuclear Physics, US DoE under Contract No. DE-AC02-98CH10886 with Brookhaven Science Associates, LLC. 
\!\!\!\!\!\!\!\!


\begin{thebibliography}{99}
\bibitem{2006Chad} M.B.~Chadwick,  P.~Oblo\v zinsk\' y, M.~Herman {\it et al.}, ``ENDF/B-VII.0: Next Generation Evaluated Nuclear Data Library for Nuclear Science and Technology," {\sc Nucl. Data Sheets} {\bf 107}, 2931 (2006).
\bibitem{2011Chad} M.B.~Chadwick, M.~Herman, P.~Oblo\v zinsk\' y {\it et al.}, ``ENDF/B-VII.1 Nuclear Data for Science and Technology: Cross Sections, Covariances, Fission Product Yields and Decay Data," {\sc Nucl. Data Sheets} {\bf 112}, 2887 (2011).
\bibitem{2010Pri} B. Pritychenko, S.F. Mughabghab, A.A. Sonzogni, ``Calculations of Maxwellian-averaged Cross Sections and Astrophysical Reaction Rates Using the ENDF/B-VII.0, JEFF-3.1, JENDL-3.3 and ENDF/B-VI.8 Evaluated Nuclear Reaction Data Libraries," {\sc Atom. Data and Nucl. Data Tables} {\bf 96}, 645 (2010).
\bibitem{2006Dil} I. Dillmann, M. Heil, F.  K\"{a}ppeler {\it et al.}, ``KADoNiS - The Karlsruhe Astrophysical Database of Nucleosynthesis in Stars,"AIP Conf. Proc. {\bf 819}, 123 (2006). Data downloaded from $\langle$http://www.kadonis.org$\rangle$ on April 7, 2010.
\bibitem{2012Pri} B. Pritychenko, S.F. Mughabghab, ``Neutron Thermal Cross Sections, Westcott Factors, Resonance Integrals, Maxwellian Averaged Cross Sections and Astrophysical Reaction Rates Calculated from the ENDF/B-VII.1, JEFF-3.1.2, JENDL-4.0, ROSFOND-2010, CENDL-3.1 and EAF-2010 Evaluated Data Libraries,"  {\sc Nucl. Data Sheets} {\bf 113}, 3120 (2012).
\bibitem{2006Mugh} S.F.~Mughabghab, ``Atlas of Neutron Resonances: Resonance Parameters and Neutron Cross Sections, Z = 1-100" (Elsevier, Amsterdam 2006).
\bibitem{2007Iaea} H.D. Choi, R.B. Firestone, R.M. Lindstrom {\it et al.}, ``Database of prompt Gamma Rays from Slow Neutron Capture for Elemental Analysis," (International Atomic Energy Agency, Vienna, 2006).
\bibitem{2008Lit} R.C. Little, T. Kawano, G.D. Hale {\it et al.}, ``Low-fidelity Covariance Project," {\sc Nucl. Data Sheets} {\bf 109}, 2828 (2008). Available from $\langle$http://www.nndc.bnl.gov/lowfi/$\rangle$.
\bibitem{2014Chad} M.B.~Chadwick, E.~Dupont, E.~Bauge {\it et al.}, ``The CIELO Collaboration: Neutron Reactions on $^{1}$H, $^{16}$O, $^{56}$Fe, $^{235,238}$U, and $^{239}$Pu," {\sc Nucl. Data Sheets} {\bf 118}, 1 (2014).
\bibitem{2002Mann} W. Mannhart, ``Response of Activation Reactions in the Neutron Field of Californium-252 Spontaneous Fission," 
{\sc International Reactor Dosimetry File 2002 (IRDF-2002)}, Technical Reports Series No. 452, (International Atomic Energy Agency, Vienna, 2006).
\bibitem{2011Kon} A. J. Koning, E. Bauge, C.J. Dean {\it et al.}, ``Status of the JEFF Nuclear Data Library," {\sc J. of the Korean Physical Society}  {\bf 59}, No. 2, 1057 (2011).
\bibitem{2011Shi} K. Shibata, T. Kawano, T. Nakagawa {\it et al.}, ``JENDL-4.0: A New Library for Nuclear Science and Engineering," {\sc J. of Nuclear Science and Technology} {\bf 48}, 1 (2011). 
\bibitem{2014Otu}  N. Otuka, V. Semkova, E.~Dupont {\it et al.}, {\sc Nucl. Data Sheets} {\bf 120}, 272 (2014). Experimental Nuclear Reaction Data (EXFOR),  http://www-nds.iaea.org/exfor.


\end{thebibliography}
\end{document}